\documentclass[conference]{IEEEtran}
\usepackage{amsmath,amssymb}
\usepackage{graphicx}
\usepackage{booktabs}
\usepackage[section]{placeins}
\usepackage{array}
\usepackage{tikz}
\usetikzlibrary{arrows.meta,positioning,calc}
\usepackage[hidelinks]{hyperref}
\hypersetup{pdfauthor={},pdfsubject={},pdfkeywords={}}

\newcommand{\on}{\mathsf{ON}}
\newcommand{\off}{\mathsf{OFF}}

\title{
Verifiable Logical-Physical Contracts for \\ Refactored On-Die Power Control Circuitry
}

\author{
\IEEEauthorblockN{Sabrina Yarzada, Paul Yi-Chia Chen, Yuanhan Li, Jeongeun Kim, Yujie Cai, and Christopher Torng}
\IEEEauthorblockA{
Department of Electrical and Computer Engineering\\
University of Southern California\\
Los Angeles, CA, USA
}
}
\begin{document}
\maketitle
\begin{abstract}
Recent literature allows independent hardware experiments to occupy a single die for scalable and cost-efficient silicon prototyping. Within this framework, specialized hardware functions including low-power control mechanisms can be offered as reusable on-die microservices. However, this requires well-defined contracts connecting digital requests to physical behavior. We present a \emph{checkable logical-physical contract formulation} including a \emph{physical contract model} describing service completion between client and provider through formally verified digital guarantees and physically validated completion conditions. We introduce \emph{Lightswitch}, a proof-of-concept on-die power control microservice. We assess post-layout designs in TSMC 28nm and demonstrate how checkable contracts provide formal guarantees.
\end{abstract}

\section{Introduction}
\label{sec-introduction}

Silicon prototyping is essential for evaluating hardware designs under realistic physical conditions, but fabrication costs and design effort make it difficult to obtain silicon feedback early and frequently~\cite{venn-tinytapeout-2024}. Traditional multi-project wafer (MPW) approaches share fabrication costs across designs~\cite{kahng-mpw-tcad2007, cohen-mosis-1981, conway-mosis-mpc-1981, tomovich-mosis-1988}, but each project still faces substantial integration, implementation overhead, and costs.

\begin{figure}[t]
\centering
\includegraphics[width=\linewidth]{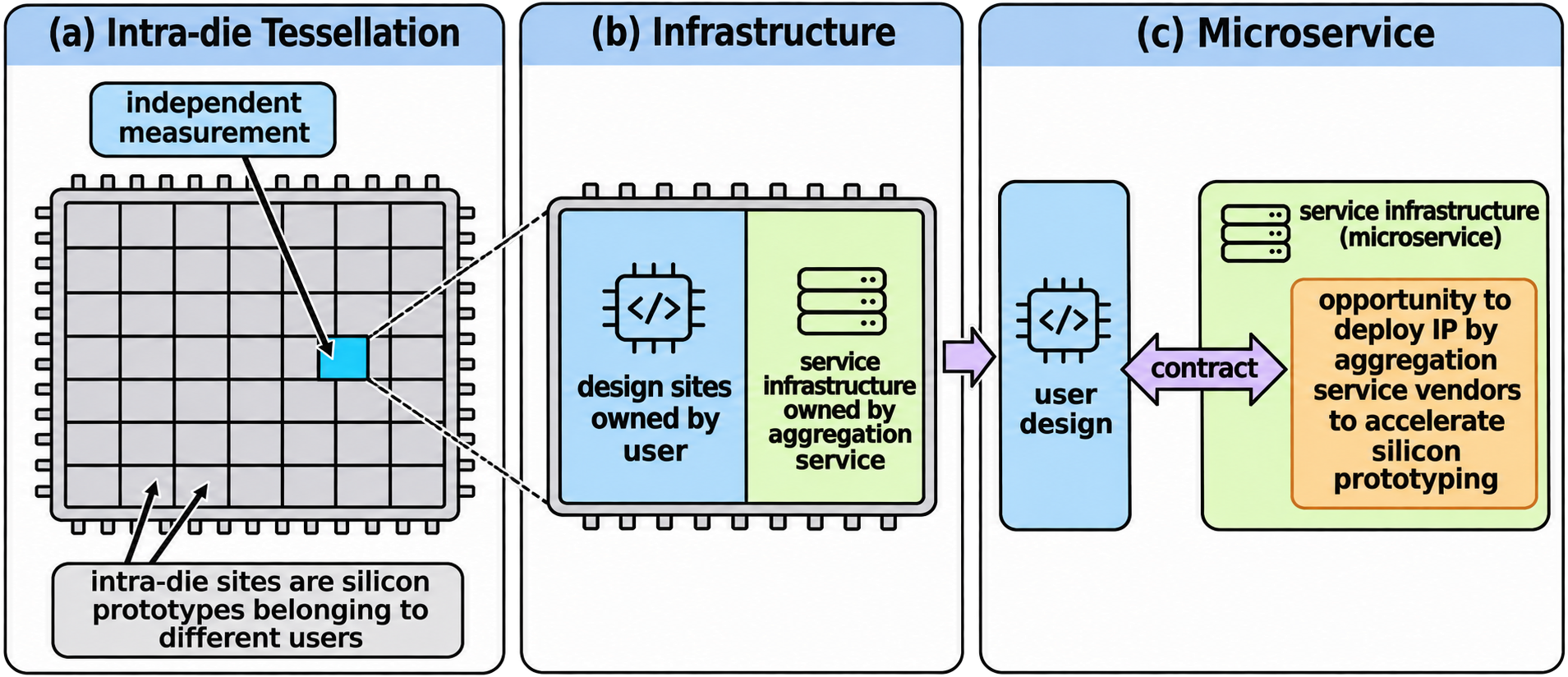}
\centering
\caption{\textbf{On-die microservices for silicon prototyping.} (a) Independent user sites share a single die. (b) User designs coexist with shared infrastructure. (c) Reusable microservices provide functionality through well-defined contracts.}
\label{fig-intro}
\end{figure}

Recent literature on silicon query architectures introduces a new capability to place independent hardware experiments alongside others within an intra-die tessellation on a single die~\cite{kim-chipstitch-2026}. Figure~\ref{fig-intro}(a) shows how final full-chip integration is handled separately by an external on-die aggregation service similar to multi-project wafer services~\cite{mosis-web, muse-semi-web}, which would be envisioned to leverage the prior work's research contributions to provide research-grade timing with high-speed multi-GHz clock distribution throughout the intra-die tessellation and practical, correct-by-construction physical integration benefits.

Beyond sharing silicon area, such architectures create an opportunity to share hardware functionality. Common services such as power control, clock management, and monitoring could be implemented once and exposed to individual experiment sites through reusable interfaces~\cite{nayak-cgra-power-2023}. As illustrated in Figure~\ref{fig-intro}(b) and~(c), user designs and shared service infrastructure are owned by separate parties. This separation requires explicit contracts defining what the client expects and what the provider guarantees~\cite{benveniste-contracts-2018,saoud-assume-guarantee-2021}. For services that interact with physical hardware, these guarantees must connect digital interface behavior to physical conditions that can be checked.

A central challenge revolves around different designs integrated by different parties on the same die and defining what constitutes physical service completion. Digital verification alone cannot establish continuous physical behavior from finite observations. Therefore, a logical--physical contract must include a \emph{physical contract model} that defines which observations are sufficient to satisfy the client's requirements. For example, this model may specify voltage thresholds, measurement locations, and observation times. The client and provider agree on this model as part of the contract, making the physical meaning of digital acknowledgment explicit. Different models can trade off physical coverage, response latency, and implementation cost. The resulting guarantees hold relative to the agreed model and its operating assumptions.

Power control illustrates this challenge. A site can request to turn ON or OFF through a digital interface, but completion depends on analog rail behavior. During power-up, the switched supply must charge through the power switches before the site can safely resume operation. This transition depends on load capacitance, parasitics, process-voltage-temperature (PVT) variation, and switching activity. A fixed waiting period may delay acknowledgment unnecessarily, while premature acknowledgment can expose the site to unsettled supply voltage. Further, enabling too many power switches simultaneously can induce inrush currents and supply disturbances that affect other sites on the die~\cite{shin-power-gating-2010,jiang-power-gating-2008,singh-powergating-2012}.

Prior work has addressed similar challenges through power-gating wake-up scheduling~\cite{lee-wakeup-tcad-2012,kim-wakeup-scheduling-2016}, ring oscillator (RO)-based supply monitoring~\cite{ogasahara-aspdac-2008,ogasahara-jssc-2009,du-voltage-sensor-2023}, and staged power switching~\cite{kim-power-gating-2003,jiang-power-gating-2008}. These techniques provide building blocks for power control, but do not establish an explicit agreement connecting digital acknowledgment to a client-accepted model of physical completion.

This work presents a \emph{checkable logical--physical contract formulation} for on-die microservices. The key insight is that the physical meaning of service completion must explicitly be part of the contract. The formulation separates the agreed physical contract model, the digital service guarantees, and the evidence connecting them. Formal verification checks that the controller follows its digital guarantees, and physical characterization validates the relationship between sensor decisions and the agreed physical model. Together, these establish a checkable connection between digital acknowledgment and physical behavior under specified operating conditions.

This work also presents \emph{Lightswitch}, a power control mechanism as a first proof-of-concept on-die microservice, which we leverage to study our contract formulation in detail. Lightswitch initially builds upon Chipstitch's~\cite{kim-chipstitch-2026} basic site-based architecture and perimeter-based power switching. Lightswitch then exposes a common request--acknowledge interface with formally checked acknowledgment, isolation, switching order, and fault handling. Physically characterized ring oscillator (RO) canaries detect ON readiness, characterized discharge timing determines OFF completion, and staged perimeter switching controls for wake-up disturbance. Configurable sensing and switching mechanisms allow the contract formulation to be studied across sites with different physical requirements.

This paper makes three contributions:
\begin{itemize}
  \item \textbf{A logical--physical contract formulation.} A checkable contract that includes a client-agreed physical contract model, connecting formally verified digital guarantees to physically validated completion conditions.

  \item \textbf{A proof-of-concept microservice.} Lightswitch implements the formulation using RO-based readiness detection, characterized OFF completion, and staged perimeter switching through a common client interface.

  \item \textbf{An evaluation of contract validation and tradeoffs.} Three differently sized TSMC 28\,nm sites are used to study contract correctness, alternative physical models, acknowledgment latency, power integrity, and implementation overhead.
\end{itemize}

To our knowledge, we are the first to develop a formal contract mechanism for on-die microservices that work to establish logical-physical contractual guarantees between different designs integrated by different parties on the same die.

\section{Logical--Physical Microservice Contract}\label{sec-contract}

\begin{figure}[t]
\centering
\includegraphics[width=\linewidth]{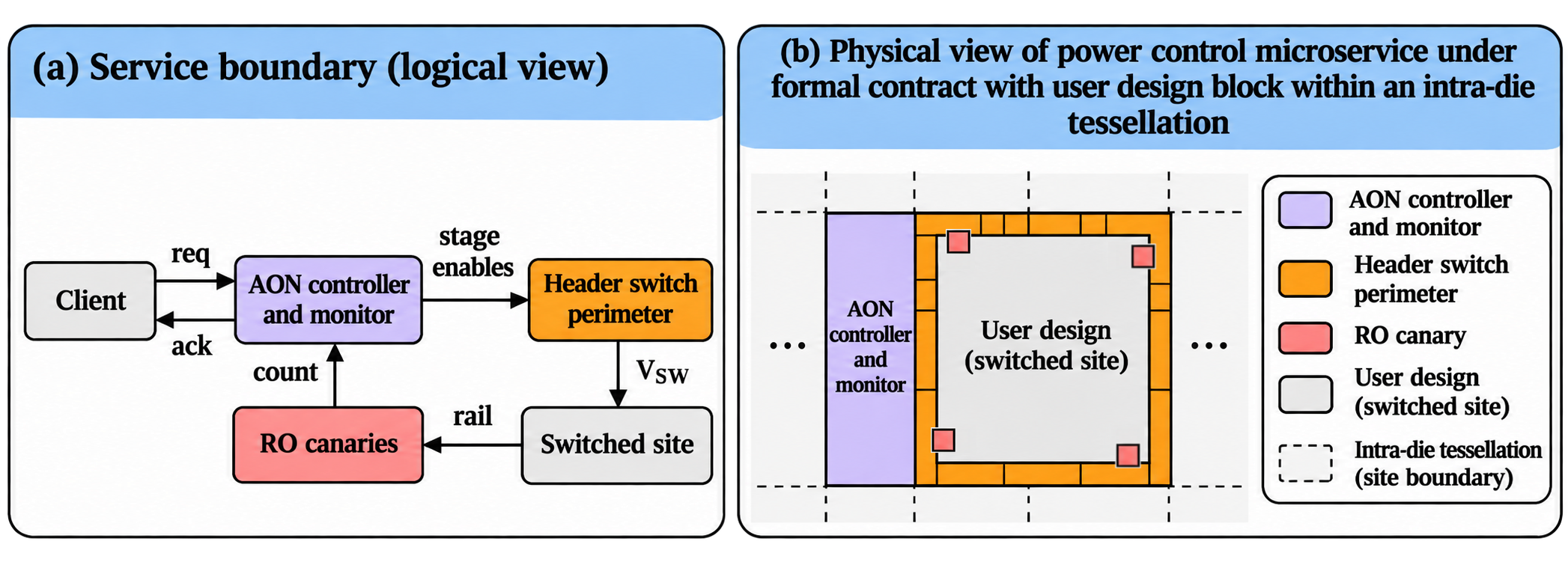}
\centering
\caption{Lightswitch architecture. (a) Logical view of the service boundary between the client and provider. (b) Physical view of the power control microservice surrounding a user design within an intra-die tessellation.}
\label{fig-service}
\end{figure}

\begin{table}[t]
\caption{Lightswitch client and provider responsibilities.}
\label{tbl-responsibilities}
\centering\footnotesize\setlength{\tabcolsep}{3pt}
\begin{tabular}{@{}p{.16\linewidth}p{.37\linewidth}p{.35\linewidth}@{}}
\toprule
Phase & Client responsibility & Provider responsibility \\
\midrule
Design &
Specify site requirements and accept the physical contract model. &
Define, validate physical model; implement switching, sensing, power intent. \\
Runtime &
Issue valid ON/OFF requests and wait for corresponding acknowledgment. &
Control transitions, maintain isolation, check readiness, report state or faults. \\
\bottomrule
\end{tabular}
\end{table}

This section introduces a logical--physical contract formulation for on-die microservices. The contract defines client and provider responsibilities, the agreed meaning of physical completion, and the evidence required to check each guarantee. Lightswitch serves as a concrete example using power control.

\subsection{Client and Provider Responsibilities}

On-die microservices separate the client's requests from the provider's responsibility to execute them. At design time, the client specifies its requirements and accepts a model defining physical service completion. The provider implements and validates the service. At run time, the client issues requests, and the provider reports completion or faults.

Figure~\ref{fig-service} shows an abstract view of Lightswitch. An experiment occupies a switched power domain. The platform has an always-on (AON) controller, perimeter header switches, and ring oscillator (RO) canaries. Through a request-acknowledge interface~\cite{sutherland-micropipelines-1989}, the client requests a power state $q\in\{\off,\on\}$ and receives acknowledgment and status. The provider manages switching, rail monitoring, and fault handling.

Table~\ref{tbl-responsibilities} summarizes the responsibilities for Lightswitch. The client and provider agree on physical completion requirements during integration. At run time, the provider checks these requirements before acknowledgment. After ON acknowledgment, the client performs the required clock and reset sequence before starting the experiment.

\subsection{Physical Contract Model and Guarantees}

A logical--physical microservice contract is defined as $\mathcal C=(A,G,\mathcal E,\mathcal M)$, where $A$ specifies client and environment assumptions, $G$ defines provider guarantees, $\mathcal E$ specifies the supported operating range, and $\mathcal M$ is the \emph{physical contract model}. The model defines the physical observations that client and provider agree are sufficient to establish service completion.

For an operation $q$, let $R_{\mathcal M}(q,t)$ denote completion under the agreed physical model. The central contract guarantee is
\begin{equation}\label{eq:contract}
 A\land\mathit{ack}_q(t)
 \;\Longrightarrow\;R_{\mathcal M}(q,t),
\end{equation}
for operating conditions within $\mathcal E$, formalizing the connection between digital acknowledgment and the physical meaning of service completion.

For Lightswitch, ON completion requires the switched supply to settle to an acceptable voltage. A given physical contract model formalizes this meaning. Let $V_{\rm SW}(x,t)$ denote the switched-rail voltage at location $x$ and time $t$. A physical model may specify observation locations $\mathcal X$, sample times $\mathcal U_t\subseteq[t-\tau,t]$, a settling interval $\tau$, and a minimum voltage $V_{\min}$, and formulate an agreed-upon ON readiness condition
\begin{equation}\label{eq:ready}
 R_{\mathcal M}^{\on}(t)=
 \bigwedge_{x\in\mathcal X}
 \bigwedge_{u\in\mathcal U_t}
 [V_{\rm SW}(x,u)\ge V_{\min}].
\end{equation}

Figure~\ref{fig-service}(b) shows an abstract view of observable locations $\mathcal X$ concretely drawn as RO canaries, although the contract does not require that they be implemented as such. The provider selects candidate locations and sample times through physical characterization, and the client agrees on the resulting model. Samples do not need to be uniformly spaced in space or time. For example, a client may accept fewer observation locations to reduce sensing overhead and response latency. Satisfying the model establishes readiness at the agreed observations.

For Lightswitch OFF acknowledgment, the agreed completion conditions require disabled switches, maintained isolation, and a characterized discharge interval. The contract also specifies power integrity limits. Let $D_{\rm AON}$ denote maximum source/AON supply droop and $S_I$ maximum supply-current slew. The agreed requirements are
\begin{equation}\label{eq:pi}
 D_{\rm AON}\le D_{\max},\qquad S_I\le S_{\max}.
\end{equation}

Response latency is measured relative to the agreed completion model. Let $c_{\rm oracle}$ denote the first controller clock edge at which the model is satisfied, and $c_{\rm ack}$ the acknowledgment edge. The additional waiting time is
\begin{equation}\label{eq:delta}
 \Delta=c_{\rm ack}-c_{\rm oracle}.
\end{equation}
A negative $\Delta$ indicates premature acknowledgment under the agreed model, and a positive $\Delta$ measures additional waiting.

A client may choose various physical contract models at design time that trade off fidelity vs.~coverage, latency, and cost, while being mathematically formulated for proof rigor.

\subsection{Checking the Digital and Physical Guarantees}

The contract is checked through two obligations. Digital verification establishes that acknowledgment follows the implementation's completion indication. Physical validation checks that this indication satisfies the agreed physical contract model under supported operating conditions.

In Lightswitch, the completion indication for ON requests is the synchronized sensor ready signal $Q_\theta$. The controller must acknowledge only when this signal is asserted and the sensing mechanism must indicate readiness only when the agreed ON completion model is satisfied:
\begin{equation}\label{eq:twolevel}
 \underbrace{\mathit{ack}_{\on}\Rightarrow Q_\theta}_{\text{RTL proof}},\qquad
 \underbrace{A\land Q_\theta\Rightarrow R_{\mathcal M}^{\on}}_{\text{physical validation}}.
\end{equation}
Both obligations apply within the supported operating range. The first is checked through RTL formal verification of acknowledgment, switch sequencing, isolation, fault handling, and recovery. The second is evaluated by comparing sensor decisions against independent rail voltage measurements at the agreed locations and times. Different RO configurations can be validated against the same physical contract model.

For OFF acknowledgment, formal verification checks that the switches are disabled, isolation is maintained, and the configured discharge interval has completed. Physical characterization validates the discharge timing under the agreed operating conditions. Together, these checks connect the digital guarantees to the accepted physical contract model.
\section{Lightswitch Architecture and Configuration}\label{sec-impl}

Lightswitch is a proof-of-concept power control microservice used to study logical-physical contract formulations. It implements the contract through staged perimeter switching and RO-based readiness detection. Sensing and switching parameters are configured to satisfy different physical contract models while preserving a common client interface.

\subsection{Staged Perimeter Switching}

Lightswitch controls power delivery through header switches placed around the perimeter of each site. These switches connect the shared supply to the site's internal rail and are divided into independently controlled segments. During power-up, the AON controller enables the segments sequentially to spread the charging current over time. This configurable power-up staging limits peak inrush current and reduces disturbances to the shared supply~\cite{jiang-power-gating-2008,lee-wakeup-tcad-2012}.

The switching sequence is described by $\pi=\{(w_k,\delta_k)\}_{k=1}^{K}$, where $w_k$ is the switch width enabled at stage $k$ and $\delta_k$ is the delay before the next stage. The number, size, and placement of header switches determine the available current capacity, and the stage schedule controls how quickly that capacity is enabled. These choices affect rail recovery time, power integrity, and implementation overhead.

Staging introduces a tradeoff between wake-up time and supply disturbance. More stages can reduce peak current but increase wake-up latency, and larger switches reduce supply resistance at the cost of area and potentially higher inrush current. Lightswitch allows these parameters to vary with each site's requirements without changing the client interface.

\subsection{Ring Oscillator-Based Readiness Detection}
\label{sec-ro}
Lightswitch uses RO canaries to detect physical rail readiness because their frequency responds to local supply voltage and their transitions can be measured digitally~\cite{ogasahara-jssc-2009}. Each RO is powered by the switched supply and an AON monitor counts its oscillations over a fixed sampling window. Distributed canaries allow readiness to be estimated at different locations across a site.

A key challenge is distinguishing voltage changes from process and temperature variation, which also affect RO frequency. Lightswitch supports two sensing methods. Single-RO sensing compares each canary's absolute count against calibrated thresholds. Dual-RO sensing normalizes these counts using a reference RO powered by the AON supply to reduce shared variation. For canary $i$ in sampling window $k$,
\begin{equation}
x_i[k]=
\begin{cases}
n_i[k], & \text{single RO},\\
n_i[k]/n_{\mathrm{ref}}[k], & \text{dual RO},
\end{cases}
\end{equation}
where $n_i[k]$ and $n_{\mathrm{ref}}[k]$ are the respective oscillation counts. Both methods require calibration because normalization does not completely eliminate PVT sensitivity.

Lightswitch qualifies the rail when every selected canary satisfies its calibrated limits for $m$ consecutive windows:
\begin{equation}\label{eq-qualify}
Q_{\theta}[k]=
\bigwedge_{i=1}^{N_{\rm RO}}
\bigwedge_{j=0}^{m-1}
[H_i^{\min}\le x_i[k-j]\le H_i^{\max}].
\end{equation}
The sampling window and $m$ balance measurement resolution, stability, and acknowledgment latency. Qualification requires new, synchronized measurements and a valid, non-zero reference count for dual-RO sensing.

\begin{figure}[t]
\centering
\begin{tikzpicture}[font=\scriptsize,>=Latex]
\foreach \dx/\nm/\ct in {0/{1 RO}/1,1.7/{2 ROs}/2,3.4/{4 ROs}/4}{
 \begin{scope}[xshift=\dx cm]
 \draw[fill=gray!6,thick] (0,0) rectangle (1.2,.78);
 \draw[line width=2pt,orange!65] (0,0) rectangle (1.2,.78);
 \node[anchor=north] at (.6,-.07) {\nm};
 \ifnum\ct=1
   \fill[blue!75] (.6,.38) circle (2.0pt);
 \fi
 \ifnum\ct=2
   \fill[blue!75] (.18,.38) circle (2.0pt);
   \fill[blue!75] (1.02,.38) circle (2.0pt);
 \fi
 \ifnum\ct=4
   \foreach \x in {.18,1.02} {
     \foreach \y in {.14,.64} {
       \fill[blue!75] (\x,\y) circle (2.0pt);
     }
   }
 \fi
 \draw[->] (-.27,.38)--(0,.38);
 \end{scope}}
\end{tikzpicture}
\caption{One, two, and four distributed RO canaries (blue) with perimeter power switches (orange). An optional AON reference RO supports normalization.}
\label{fig-config}
\end{figure}
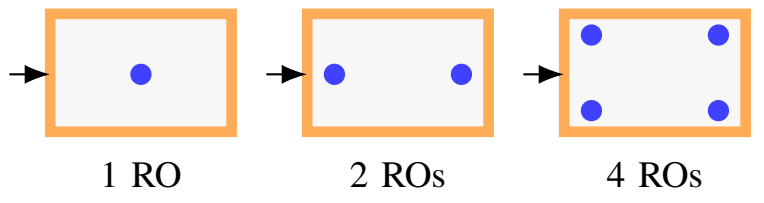

The number and placement of RO canaries can be adjusted to meet different physical contract models. Figure~\ref{fig-config} illustrates the one-, two-, and four-canary configurations evaluated in Section~\ref{sec-eval}. Dual-RO sensing adds one shared AON reference RO to each configuration. These configurations provide different sensing coverage and overhead, but must be validated against the selected physical model before use.

\subsection{Site-Specific Configuration}

Lightswitch demonstrates how different implementations can satisfy a common logical-physical contract formulation. For an agreed physical contract model $\mathcal M$, the provider selects a configuration $z=(W,\pi,E,C,\theta)$, where $W$ is the total switch capacity, $\pi$ the stage schedule, $E$ the supply connection placement, $C$ the number and locations of RO canaries, and $\theta$ the sensing parameters. These choices affect acknowledgment latency, supply disturbance, and implementation overhead.

To configure a site, the client and provider first agree on the physical completion requirements and supported operating conditions. The provider then selects candidate sensing parameters from pre-characterized configurations, including the sensing method, sampling window, and thresholds. These are combined with site-specific switching and placement choices. Independent rail voltage measurements determine whether the resulting configuration satisfies the agreed physical model across characterized conditions. Configurations that fail validation are adjusted or rejected.

The provider selects a configuration that minimizes acknowledgment latency subject to agreed contract requirements:
\begin{equation}\label{eq-choose}
 \min_z T_{\rm ack}(z)\quad\text{s.t.}\quad
 \begin{cases}
 \Delta(z,e)\ge 0, & e\in\mathcal E,\\
 D_{\rm AON}(z,e)\le D_{\max}, & e\in\mathcal E,\\
 S_I(z,e)\le S_{\max}, & e\in\mathcal E,\\
 A_{\rm area}(z)\le A_{\max}. &
 \end{cases}
\end{equation}
Here, $e$ represents an operating condition within the range $\mathcal E$, and $\Delta$ is measured relative to readiness under the agreed physical model $\mathcal M$. Configurations that acknowledge prematurely or violate power integrity and area limits are rejected.

Switch placement and RO locations are fixed during physical implementation. Stage timing and sensing thresholds may be programmable if supported by the hardware. The selected parameters are stored in the AON controller. Thus, different sites can therefore use different physical models and implementations while preserving the same digital interface and contract verification method.

\section{Evaluation}\label{sec-eval}

This section evaluates the proposed logical-physical contract formulation using Lightswitch as a proof-of-concept microservice. It first checks the digital and physical contract guarantees, then studies how different physical contract models affect acknowledgment latency, sensing requirements, and implementation overhead. 

\subsection{Evaluation Methodology}
\label{sec-eval-setup}

We build three variants of TSMC 28\,nm post-layout client designs, each with a digital power control state machine coupled to a synthetic load (e.g., FO4-loaded inverters, NAND, and NOR gates connected in a loop), with distinct load characteristics and physical floorplans (150$\times$150um, 175$\times$175um, and 350$\times$350um). All sites share the same digital interface and contract formulation, with site-specific physical contract models and implementation configurations. Yosys verifies the digital controller using bounded model checking and induction. Spectre characterizes RO behavior and rail readiness using physical designs implemented in Cadence Innovus. An independent voltage oracle validates sensor decisions and acknowledgment timing against the agreed physical contract models across PVT and load conditions. Voltus measures wake-up power integrity, and area and power analysis quantifies the overhead of implementing and checking the contract.

\subsection{Digital Contract Verification}
\label{sec-eval-formal}

The first experiment verifies the digital guarantees of the logical-physical contract. The Lightswitch controller receives ON/OFF requests, enables power switches in stages, checks rail readiness, and acknowledges completed transitions or reports faults. These behaviors are checked against the controller's RTL using Yosys SAT~\cite{wolf-yosys-2013} and SystemVerilog assertions. Table~\ref{tbl-properties} summarizes the verified guarantees.

\begin{table}[t]
\caption{Checkable digital service contract guarantees.}
\label{tbl-properties}
\centering
\footnotesize
\setlength{\tabcolsep}{3pt}
\begin{tabular}{@{}p{.21\linewidth}>{\raggedright\arraybackslash}p{.71\linewidth}@{}}
\toprule
Property & Checked guarantee \\
\midrule
ACK validity &
ACK matches the requested state and is issued only
when the controller is idle and fault-free. \\
ON ACK &
Rail-ready signal asserted, switches enabled,
isolation removed, and reset release allowed. \\
OFF ACK &
Discharge waiting period completed, switches disabled,
site isolated, and reset held. \\
Switch order &
Isolation and reset maintained during transitions;
OFF isolates the site before disabling switches. \\
Fault handling &
Timeout or loss of readiness triggers isolation
and prevents acknowledgment. \\
Response bound &
Accepted requests complete or reach a fault
within configured time limits. \\
\bottomrule
\end{tabular}
\end{table}

The verification covers three controller configurations with different numbers of independently controlled power switch groups and OFF discharge delays. The discharge delay is the number of clock cycles the controller waits after disabling the switches to allow the site's supply voltage to fall. The configurations test normal operation, a larger controller, and an invalid zero delay setting that must prevent OFF acknowledgment. Each is checked using 40-cycle bounded model checking~\cite{biere-bmc-1999} and induction up to 32 steps~\cite{sheeran-induction-2000,een-induction-2003}. The verification uses the original controller RTL and checks its behavior under arbitrary client requests and sensor inputs.

\begin{table}[t]
\caption{Digital contract verification results.}
\label{tbl-formal}
\centering
\footnotesize
\setlength{\tabcolsep}{3pt}
\begin{tabular*}{\columnwidth}{@{\extracolsep{\fill}}lcccc@{}}
\toprule
Configuration & Groups & OFF delay & BMC / Ind. & Witnesses \\
 & (\#) & (cycles) & & \\
\midrule
Standard controller & 2 & 4 & PASS / PASS & 6/6 \\
Zero delay fault & 2 & 0 & PASS / PASS & 1/1 \\
Expanded controller & 4 & 12 & PASS / PASS & 2/2 \\
\bottomrule
\end{tabular*}
\end{table}

As shown in Table~\ref{tbl-formal}, all three configurations pass both verification checks, for six passing checks in total. Separate reachability tests find nine examples covering normal transitions, sensor waiting, readiness loss, fault recovery, and watchdog expiration. In particular, the zero delay configuration correctly refuses OFF acknowledgment because no discharge waiting period is provided. These results establish that the controller satisfies the digital contract guarantees, including acknowledgment, isolation, and fault-handling.

\subsection{RO Calibration and Readiness Validation}\label{sec-eval-ro}
\begin{figure}[t]
\centering
\includegraphics[width=\linewidth]{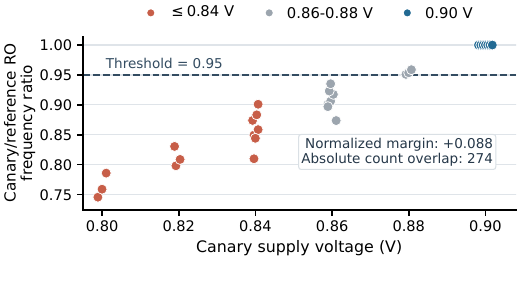}
\centering
\caption{RO sensing across 30 voltage and PVT conditions. Reference normalization distinguishes low and nominal voltages.}
\label{fig-ro-calib}
\end{figure}
Section~\ref{sec-ro} introduced two RO sensing methods for detecting rail readiness. This experiment evaluates how well each method supports an agreed physical contract model across PVT conditions. Pre-layout Spectre simulations compare absolute RO counts and reference-normalized counts across 30 voltage and PVT conditions. The reference RO operates at 0.90\,V, and the canary voltage ranges from 0.80 to 0.90\,V. RO frequency increases with supply voltage, and reference normalization helps reduce PVT effects. The simulated frequencies are used to estimate counts over a 512-cycle sampling window. The evaluated model distinguishes low voltage ($V_{\rm SW}\leq0.84$\,V) from nominal voltage ($V_{\rm SW}=0.90$\,V).

Figure~\ref{fig-ro-calib} shows the normalized RO frequency ratios. Absolute counting cannot reliably distinguish low and nominal voltages across PVT because their estimated count ranges overlap by 274 counts. Reference normalization separates the two voltage classes with a margin of 0.088. A threshold of 0.95 separates all 13 low-voltage and seven nominal-voltage intervals in the sampled data.

These results show how the choice of sensing method affects the physical contract models that a microservice can support. Reference normalization provides consistent voltage classification across the sampled PVT conditions, allowing the same threshold to be used without corner-specific calibration. This demonstrates how physical characterization can guide the selection of a sensing implementation that meets the agreed contract requirements.

\subsection{Physical Contract Model and Canary Sensitivity}
\label{sec-eval-contract}
\begin{figure}[t]
\centering
\includegraphics[width=\linewidth]{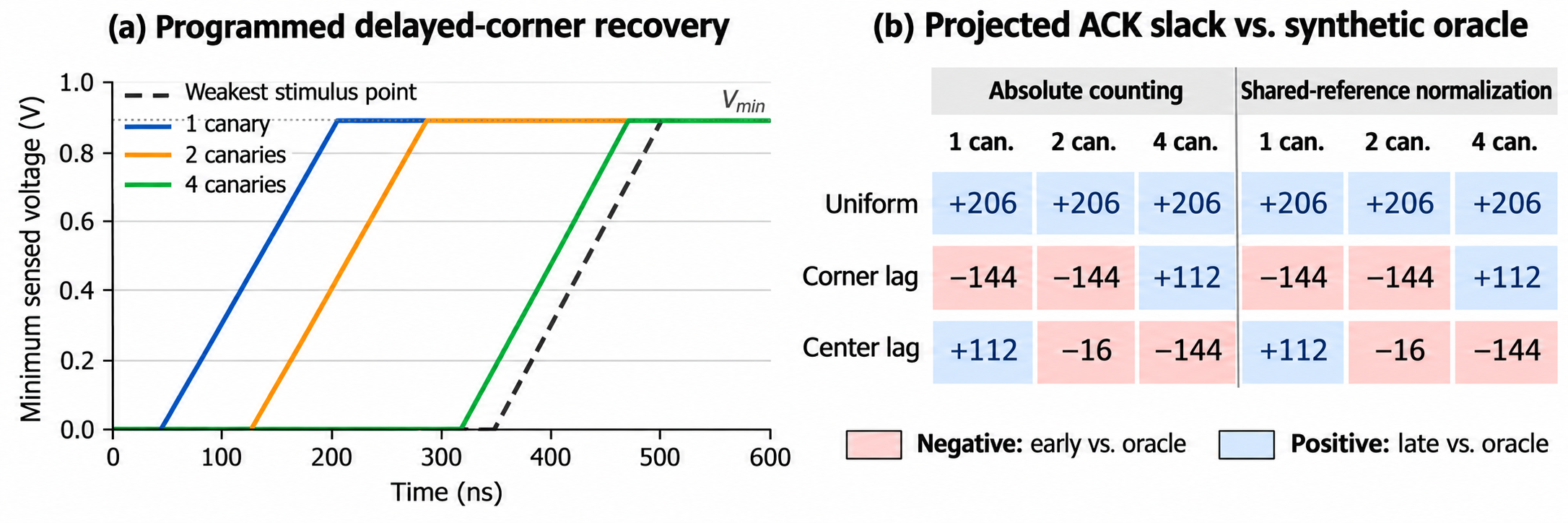}
\centering
\caption{Canary placement and projected acknowledgment timing under synthetic rail recovery. (a) Delayed-ramp site corner recovery. (b) Acknowledgment slack across three recovery scenarios. Negative values indicate early acknowledgment relative to the synthetic oracle.}
\label{fig-eval-canaries}
\end{figure}

Beyond selecting a sensing method, the number and placement of RO canaries also affect how well Lightswitch can meet the agreed physical contract model. To study these effects, the one-, two-, and four-canary configurations in Fig.~\ref{fig-config} are evaluated using synthetic rail-voltage traces representing uniform recovery, a delayed-ramp site corner, and a delayed-ramp site center. Previously characterized RO frequencies are used to estimate acknowledgment times with 128-cycle sampling windows. These are compared against a voltage oracle that tracks the slowest ramped rail location.

Figure~\ref{fig-eval-canaries} shows how canary placement affects acknowledgment timing. Under delayed-ramp site corner recovery, one and two canary configurations acknowledge 144 cycles before the oracle, compared with 112 cycles after readiness for four canaries, exposing a functional integrity violation. Under delayed-ramp site center recovery, the centered one-canary configuration acknowledges 112 cycles after readiness, but four corner canaries acknowledge 144 cycles early. All configurations have 206 cycles of additional waiting under uniform recovery.

These results show that readiness detection depends on canary placement, not just the number of canaries. The physical contract model allows the client and provider to agree on which locations must be monitored to establish readiness. This makes it possible to choose sensing configurations that balance physical coverage, acknowledgment latency, and implementation overhead based on the client's requirements.

\subsection{Wake-Up Power Integrity and Physical Validation}
\begin{figure}[t]
\centering
\includegraphics[width=\linewidth]{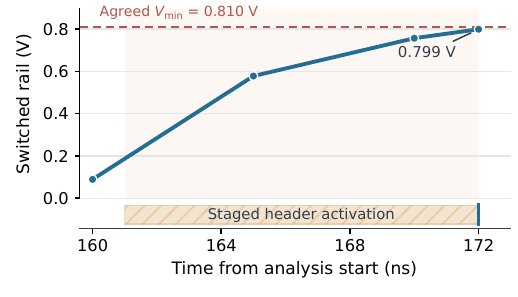}
\centering
\caption{Staged wake-up behavior of a TSMC 28\,nm systolic array site.
All 236 perimeter header switches are enabled by 172\,ns, alongside the switched-rail voltage response.
}
\label{fig-power-up}
\end{figure}

Beyond verifying digital control and calibrating RO sensing, the physical contract also requires the switched supply to reach the agreed readiness condition. To evaluate this, Lightswitch is implemented in a placed-and-routed TSMC 28\,nm systolic array site with 236 perimeter header switches. Gate-level waveforms confirm staged activation between approximately 161 and 172\,ns, with all switches enabled by 172\,ns. Voltus reports an 11.05\,mA peak inrush current during this transition.

Figure~\ref{fig-power-up} shows the initial switched rail voltage response. As the header switches turn on, the rail rises from approximately 0.09\,V to 0.58\,V at 165\,ns, reaches 0.76\,V at 170\,ns, and reaches 0.799\,V near 172\,ns. These measurements characterize the initial charging behavior of the physical power network and provide electrical evidence to complement the formally verified digital switching sequence.

\subsection{Microservice Reuse and Contract Overhead}
\label{sec-eval-reuse}

The final experiment measures the implementation cost of supporting the logical-physical contract across sites with different physical requirements. Lightswitch preserves the same digital interface and contract verification method, with site-specific switching and sensing configurations. The evaluation covers 20 power domains across three site sizes. Table~\ref{tab:sites} summarizes their physical characteristics and area overhead.

\begin{table}[t]
\caption{Site characteristics and Lightswitch overhead.}
\label{tab:sites}
\centering
\footnotesize
\setlength{\tabcolsep}{3pt}
\begin{tabular*}{\columnwidth}{@{\extracolsep{\fill}}lccccc@{}}
\toprule
Site & Area & Perim. & Avg. headers & Added area & O/H \\
     & (mm$^2$) & ($\mu$m) & (\#) & (mm$^2$) & (\%) \\
\midrule
Small  & 0.02250  & 600  & 365.5 & 0.011810 & 52.49 \\
Medium & 0.030625 & 700  & 583.6 & 0.010926 & 35.68 \\
Large  & 0.12250  & 1400 & 1131  & 0.020026 & 16.35 \\
\bottomrule
\end{tabular*}
\end{table}

Across the 20 domains, Lightswitch uses 15,290 header switches. Added area includes the surrounding region and the Lightswitch implementation, consisting of the controller, RO canaries, AON monitoring logic, and any additional switching or isolation resources. Small sites extend 16\,$\mu$m on each side, while medium and large sites extend 16\,$\mu$m horizontally and 10\,$\mu$m vertically. The total area overhead decreases from 52.49\% for small sites to 16.35\% for large sites, showing that the relative integration cost decreases as site size increases.

These results show the total area overhead of integrating Lightswitch across different sites. The additional cost of contract checking depends on the agreed physical model and the sensing hardware needed to satisfy it. Beyond basic power switching and readiness detection, stricter models may require additional RO canaries within the user site and more counting and qualification logic in the shared AON infrastructure. These additions increase area and power consumption, and longer sampling or settling requirements can increase acknowledgment latency. Thus, contract checking introduces costs for both the client and provider, with tradeoffs between physical coverage, response time, and hardware overhead.

\section{Related Work}\label{sec-related}

Shared silicon prototyping platforms, including multi-project wafers~\cite{kahng-mpw-ispd2004,kahng-mpw-tcad2007}, Tiny Tapeout~\cite{venn-tiny-tapeout-2024}, and Chipstitch~\cite{kim-chipstitch-2026}, have made silicon experimentation more accessible. Chipstitch provides the shared infrastructure and per-site power control that motivate Lightswitch. Exposing shared hardware functions as reusable services draws on software microservices~\cite{lewis-microservices-2014,dragoni-microservices-2017} and contract-based methods for specifying and verifying component guarantees~\cite{dealfaro-interfaces-2001,sangiovanni-contracts-2012,benveniste-contracts-2018,saoud-assume-guarantee-2021,cimatti-ocra-2013,cimatti-refinement-2015}. This work builds on these methods by explicitly including a client-agreed physical completion model in the service contract and connecting it to digital acknowledgment. In low power hardware, IEEE 1801 defines power intent~\cite{ieee-std-2024}, and formal verification techniques support checking digital control behavior~\cite{biere-bmc-1999,sheeran-induction-2000,een-induction-2003,bradley-ic3-2011}. Prior work also addresses power gating and staged wake-up~\cite{mutoh-mt-cmos-1995,powell-gated-vdd-2000,calhoun-distributed-mtcmos-2004,abdollahi-power-transition-2005,jiang-power-gating-2008,shin-power-gating-2010,lee-wakeup-tcad-2012}, power grid integrity~\cite{kouroussis-grid-2003,popovich-power-grid-2007}, and RO-based supply monitoring~\cite{ogasahara-aspdac-2008,ogasahara-jssc-2009,an-ro-process-2016,du-voltage-sensor-2023}. Lightswitch uses these techniques as a proof-of-concept implementation for studying how logical-physical contracts can be defined, verified, and configured for different site requirements.

\section{Conclusion}\label{sec-conclusion}

This work presented a \emph{checkable logical--physical contract formulation} for reusable on-die microservices. The formulation includes a client-agreed physical contract model that defines service completion and connects formally verified digital guarantees to physically validated conditions. Lightswitch demonstrates this approach through a power control microservice with RO-based readiness detection and staged perimeter switching. Evaluation across three different TSMC 28\,nm sites examines contract correctness, alternative physical models, acknowledgment latency, power integrity, and the overhead of contract checking. These results demonstrate how different physical requirements can be supported through a common interface and verification method.

\bibliographystyle{IEEEtran}
\bibliography{main}

@string{JAN = {Jan}}

@string{FEB = {Feb}}

@string{MAR = {Mar}}

@string{APR = {Apr}}

@string{MAY = {May}}

@string{JUN = {Jun}}

@string{OCT = {Oct}}

@string{TCAD     = {IEEE Trans. on Computer-Aided Design of Integrated
                     Circuits and Systems (TCAD)}}

@string{COMPUTER = {IEEE Computer}}

@string{TODAES   = {ACM Trans. on Design Automation
                     of Electronic Systems (TODAES)}}

@string{DAC      = {Design Automation Conf. (DAC)}}

@string{ASPDAC   = {Asia and South Pacific Design Automation
                     Conference (ASP-DAC)}}

@string{ICCAD    = {Int'l Conf. on Computer-Aided Design (ICCAD)}}

@string{JSSC     = {IEEE Journal of Solid-State Circuits (JSSC)}}

@string{TCASI    = {IEEE Trans. on Circuits and Systems I (TCAS I)}}

@string{TVLSI    = {IEEE Trans. on Very Large-Scale Integration Systems (TVLSI)}}

@string{ISLPED   = {Int'l Symp. on Low-Power Electronics and
                     Design (ISLPED)}}

@string{CACM     = {Communications of the ACM}}

@string{SCP      = {Science of Computer Programming}}

@string{ISPD       = {Int'l Symp. on Physical Design (ISPD)}}

@string{AUSTROCHIP = {Austrochip Workshop (Austrochip)}}

@string{TACAS      = {Tools and Algorithms for the Construction and Analysis of Systems (TACAS)}}

@string{FMCAD      = {Formal Methods in Computer-Aided Design (FMCAD)}}

@string{VMCAI      = {Verification, Model Checking, and Abstract Interpretation (VMCAI)}}

@string{ESECFSE    = {European Software Engineering Conf. and Symp. on the Foundations of Software Engineering (ESEC/FSE)}}

@string{ASE        = {Int'l Conf. on Automated Software Engineering (ASE)}}

@article{kahng-mpw-ispd2004,
  title        = {Multi-project reticle floorplanning and wafer dicing},
  author       = {Andrew B. Kahng and Ion I. Mandoiu and Qinke Wang and Xu Xu and Alexander Z. Zelikovsky},
  journal      = ISPD,
  year         = {2004},
  doi          = {10.1145/981066.981082},
}

@article{kahng-mpw-tcad2007,
  title        = {Enhanced Design Flow and Optimizations for Multiproject Wafers},
  author       = {Kahng, Andrew B. and Mandoiu, Ion I. and Xu, Xu and Zelikovsky, Alexander Z.},
  journal      = TCAD,
  month        = FEB,
  year         = {2007},
}

@article{venn-tiny-tapeout-2024,
  title        = {Tiny Tapeout: A shared silicon tape out platform accessible to everyone},
  author       = {Venn, Matt},
  journal      = {IEEE Solid-State Circuits Magazine},
  month        = MAY,
  year         = {2024},
}

@article{kim-chipstitch-2026,
  title        = {{Chipstitch}: A Scalable and Cost-Efficient Platform for Accelerated Silicon Prototyping},
  author       = {Jeongeun Kim and Sabrina Yarzada and Paul Yi-Chia Chen and Yuanhan Li and Wenbo Zhu and Christopher Torng},
  journal      = ICCAD,
  year         = {2026},
  doi          = {10.1145/3831252.3833953},
  url          = {https://doi.org/10.1145/3831252.3833953},
}

@article{cohen-mosis-1981,
  title     = {MOSIS: The ARPA Silicon Broker},
  author    = {Danny Cohen and George Lewicki},
  journal   = {Proceedings of the Caltech Conference on VLSI},
  month     = JAN,
  year      = {1981},
}

@article{conway-mosis-mpc-1981,
  title     = {The MPC Adventures: Experiences with the Generation of VLSI Design and Implementation Methodologies},
  author    = {Lynn Conway},
  journal   = {Proceedings of the Caltech Conference on VLSI},
  month     = JAN,
  year      = {1981},
}

@article{tomovich-mosis-1988,
  title     = {MOSIS - A Gateway to Silicon},
  author    = {Christine Tomovich},
  journal   = {IEEE Circuits and Devices Magazine},
  month     = MAR,
  year      = {1988},
}

@misc{mosis-web,
  title     = {The MOSIS Service},
  key       = {The MOSIS Service},
  howpublished = {Online Webpage},
  note      = {\url{https://themosisservice.com/about}},
  year      = {2025 (accessed Apr 1, 2025)},
}

@misc{muse-semi-web,
  title     = {Muse Semiconductor},
  key       = {Muse Semiconductor},
  howpublished = {Online Webpage},
  note      = {\url{https://www.musesemi.com}},
  year      = {2025 (accessed Apr 1, 2025)},
}

@article{benveniste-contracts-2018,
  title        = {Contracts for System Design},
  author       = {Albert Benveniste and Beno{\^i}t Caillaud and Dejan Nickovic and Roberto Passerone and Jean-Baptiste Raclet and Philipp Reinkemeier and Alberto L. Sangiovanni-Vincentelli and Werner Damm and Thomas A. Henzinger and Kim G. Larsen},
  journal      = {Foundations and Trends in Electronic Design Automation},
  month        = MAR,
  year         = {2018},
}

@article{saoud-assume-guarantee-2021,
  title        = {Assume-guarantee contracts for continuous-time systems},
  author       = {Adnane Saoud and Antoine Girard and Laurent Fribourg},
  journal      = {Automatica},
  month        = JUN,
  year         = {2021},
}

@article{sangiovanni-contracts-2012,
  title        = {Taming Dr. Frankenstein: Contract-Based Design for Cyber-Physical Systems},
  author       = {Alberto L. Sangiovanni-Vincentelli and Werner Damm and Roberto Passerone},
  journal      = {European Journal of Control},
  year         = {2012},
  volume       = {18},
  pages        = {217--238},
  doi          = {10.3166/ejc.18.217-238},
}

@article{dealfaro-interfaces-2001,
  title        = {Interface Automata},
  author       = {Luca de Alfaro and Thomas A. Henzinger},
  journal      = ESECFSE,
  year         = {2001},
  doi          = {10.1145/503209.503226},
}

@article{cimatti-ocra-2013,
  title        = {{OCRA}: A Tool for Checking the Refinement of Temporal Contracts},
  author       = {Alessandro Cimatti and Michele Dorigatti and Stefano Tonetta},
  journal      = ASE,
  year         = {2013},
  doi          = {10.1109/ASE.2013.6693137},
}

@article{cimatti-refinement-2015,
  title        = {Contracts-Refinement Proof System for Component-Based Embedded Systems},
  author       = {Alessandro Cimatti and Stefano Tonetta},
  journal      = SCP,
  year         = {2015},
  volume       = {97},
  pages        = {333--348},
  doi          = {10.1016/j.scico.2014.06.011},
}

@article{sutherland-micropipelines-1989,
  title        = {Micropipelines},
  author       = {Ivan E. Sutherland},
  journal      = CACM,
  year         = {1989},
  volume       = {32},
  number       = {6},
  pages        = {720--738},
  doi          = {10.1145/63526.63532},
}

@misc{lewis-microservices-2014,
  title        = {Microservices},
  author       = {James Lewis and Martin Fowler},
  howpublished = {martinfowler.com},
  year         = {2014},
  url          = {https://martinfowler.com/articles/microservices.html},
}

@incollection{dragoni-microservices-2017,
  title        = {Microservices: Yesterday, Today, and Tomorrow},
  author       = {Nicola Dragoni and Saverio Giallorenzo and Alberto Lluch-Lafuente and Manuel Mazzara and Fabrizio Montesi and Ruslan Mustafin and Larisa Safina},
  booktitle    = {Present and Ulterior Software Engineering},
  publisher    = {Springer},
  year         = {2017},
  pages        = {195--216},
  doi          = {10.1007/978-3-319-67425-4_12},
}

@article{wolf-yosys-2013,
  title        = {{Yosys} -- A Free Verilog Synthesis Suite},
  author       = {Clifford Wolf and Johann Glaser},
  journal      = AUSTROCHIP,
  year         = {2013},
}

@article{biere-bmc-1999,
  title        = {Symbolic Model Checking without {BDDs}},
  author       = {Armin Biere and Alessandro Cimatti and Edmund M. Clarke and Yunshan Zhu},
  journal      = TACAS,
  year         = {1999},
  doi          = {10.1007/3-540-49059-0_14},
}

@article{sheeran-induction-2000,
  title        = {Checking Safety Properties Using Induction and a {SAT}-Solver},
  author       = {Mary Sheeran and Satnam Singh and Gunnar St{\aa}lmarck},
  journal      = FMCAD,
  year         = {2000},
  doi          = {10.1007/3-540-40922-X_8},
}

@article{een-induction-2003,
  title        = {Temporal Induction by Incremental {SAT} Solving},
  author       = {Niklas E{\'e}n and Niklas S{\"o}rensson},
  journal      = {Electronic Notes in Theoretical Computer Science},
  year         = {2003},
  volume       = {89},
  number       = {4},
  pages        = {543--560},
  doi          = {10.1016/S1571-0661(05)82542-3},
}

@article{bradley-ic3-2011,
  title        = {{SAT}-Based Model Checking Without Unrolling},
  author       = {Aaron R. Bradley},
  journal      = VMCAI,
  year         = {2011},
  doi          = {10.1007/978-3-642-18275-4_7},
}

@article{mutoh-mt-cmos-1995,
  title        = {1-{V} Power Supply High-Speed Digital Circuit Technology with Multithreshold-Voltage {CMOS}},
  author       = {Shin'ichiro Mutoh and Takakuni Douseki and Yasuyuki Matsuya and Takahiro Aoki and Satoshi Shigematsu and Junzo Yamada},
  journal      = JSSC,
  year         = {1995},
  volume       = {30},
  number       = {8},
  pages        = {847--854},
  doi          = {10.1109/4.400426},
}

@article{powell-gated-vdd-2000,
  title        = {Gated-{Vdd}: A Circuit Technique to Reduce Leakage in Deep-Submicron Cache Memories},
  author       = {Michael Powell and Se-Hyun Yang and Babak Falsafi and Kaushik Roy and T. N. Vijaykumar},
  journal      = ISLPED,
  year         = {2000},
}

@article{shin-power-gating-2010,
  title        = {Power gating: Circuits, design methodologies, and best practice for standard-cell VLSI designs},
  author       = {Shin, Youngsoo and Seomun, Jun and Choi, Kyu-Myung and Sakurai, Takayasu},
  journal      = TODAES,
  month        = OCT,
  year         = {2010},
}

@article{kim-power-gating-2003,
  title        = {Understanding and minimizing ground bounce during mode transition of power gating structures},
  author       = {Suhwan Kim and Stephen V. Kosonocky and Daniel R. Knebel},
  journal      = ISLPED,
  year         = {2003},
  doi          = {10.1145/871506.871515},
}

@article{jiang-power-gating-2008,
  title        = {Power gating scheduling for power/ground noise reduction},
  author       = {Jiang, Hailin and Marek-Sadowska, Malgorzata},
  journal      = DAC,
  year         = {2008},
}

@article{lee-wakeup-tcad-2012,
  title        = {Efficient Wakeup Scheduling Considering Both Resource Usage and Timing Budget for Power Gating Designs},
  author       = {Ming-Chao Lee and Yiyu Shi and Shih-Chieh Chang},
  journal      = TCAD,
  month        = JUN,
  year         = {2012},
}

@article{abdollahi-power-transition-2005,
  title        = {An Effective Power Mode Transition Technique in {MTCMOS} Circuits},
  author       = {Afshin Abdollahi and Farzan Fallah and Massoud Pedram},
  journal      = DAC,
  year         = {2005},
  doi          = {10.1145/1065579.1065594},
}

@article{calhoun-distributed-mtcmos-2004,
  title        = {A Leakage Reduction Methodology for Distributed {MTCMOS}},
  author       = {Benton H. Calhoun and Frank A. Honor{\'e} and Anantha P. Chandrakasan},
  journal      = JSSC,
  year         = {2004},
  volume       = {39},
  number       = {5},
  pages        = {818--826},
  doi          = {10.1109/JSSC.2004.826335},
}

@book{popovich-power-grid-2007,
  title        = {Power Distribution Networks with On-Chip Decoupling Capacitors},
  author       = {Mikhail Popovich and Andrey V. Mezhiba and Eby G. Friedman},
  publisher    = {Springer},
  year         = {2007},
  doi          = {10.1007/978-0-387-71601-5},
}

@article{kouroussis-grid-2003,
  title        = {A Static Pattern-Independent Technique for Power Grid Voltage Integrity Verification},
  author       = {Dionysios Kouroussis and Farid N. Najm},
  journal      = DAC,
  year         = {2003},
}

@misc{ieee-std-2024,
  title        = {{IEEE Std 1801-2024}: IEEE Standard for Design and Verification of Low-Power Energy-Aware Electronic Systems},
  author       = {{IEEE}},
  howpublished = {IEEE Standard},
  year         = {2024},
}

@article{ogasahara-aspdac-2008,
  title        = {Dynamic supply noise measurement circuit composed of standard cells suitable for in-site {SoC} power integrity verification},
  author       = {Yasuhiro Ogasahara and Masanori Hashimoto and Takao Onoye},
  journal      = ASPDAC,
  year         = {2008},
}

@article{ogasahara-jssc-2009,
  title        = {All-Digital Ring-Oscillator-Based Macro for Sensing Dynamic Supply Noise Waveform},
  author       = {Yasuhiro Ogasahara and Masanori Hashimoto and Takao Onoye},
  journal      = JSSC,
  month        = MAY,
  year         = {2009},
}

@article{du-voltage-sensor-2023,
  title        = {An All-Digital, 1.92--7.32 mV/LSB, 0.5--2 GS/s Sample Rate, and 0-Latency Prediction Voltage Sensor With Dynamic {PVT} Calibration for Droop Detection and {AVS} System},
  author       = {Yuxuan Du and Junyi Qian and Zhuo Chen and Weiwei Shan},
  journal      = TCASI,
  year         = {2023},
  volume       = {70},
  number       = {12},
  pages        = {4888--4899},
  doi          = {10.1109/TCSI.2023.3305038},
}

@article{an-ro-process-2016,
  title        = {All-Digital On-Chip Process Sensor Using Ratioed Inverter-Based Ring Oscillator},
  author       = {Young-Jae An and Dong-Hoon Jung and Kyungho Ryu and Hyuck Sang Yim and Seong-Ook Jung},
  journal      = TVLSI,
  year         = {2016},
  volume       = {24},
  number       = {11},
  pages        = {3232--3242},
  doi          = {10.1109/TVLSI.2016.2550603},
}

@article{venn-tinytapeout-2024,
  author  = {Matt Venn},
  title   = {{Tiny Tapeout}: A Shared Silicon Tapeout
             Platform Accessible to Everyone},
  journal = {IEEE Solid-State Circuits Magazine},
  volume  = {16},
  number  = {2},
  pages   = {20--29},
  year    = {2024}
}

@article{nayak-cgra-power-2023,
  author  = {Ankita Nayak and Keyi Zhang and
             Rajsekhar Setaluri and Alex Carsello and
             Makai Mann and Christopher Torng and
             Stephen Richardson and Rick Bahr and
             Pat Hanrahan and Mark Horowitz and
             Priyanka Raina},
  title   = {Improving Energy Efficiency of {CGRAs}
             with Low-Overhead Fine-Grained Power Domains},
  journal = {ACM Transactions on Reconfigurable
             Technology and Systems},
  volume  = {16},
  number  = {2},
  year    = {2023},
  doi     = {10.1145/3558394}
}

@article{kim-wakeup-scheduling-2016,
  author  = {Sangmin Kim and Seungwhun Paik and
             Seokhyeong Kang and Youngsoo Shin},
  title   = {Wakeup Scheduling and Its Buffered Tree
             Synthesis for Power Gating Circuits},
  journal = {Integration, the VLSI Journal},
  volume  = {53},
  pages   = {157--170},
  year    = {2016},
  doi     = {10.1016/j.vlsi.2015.12.008}
}

@article{singh-powergating-2012,
  author  = {Rahul Singh and Jong Kwan Woo and
             Hyunjoong Lee and So Young Kim and Suhwan Kim},
  title   = {Power-Gating Noise Minimization by
             Three-Step Wake-Up Partitioning},
  journal = TCASI,
  volume  = {59},
  number  = {4},
  pages   = {749--762},
  year    = {2012},
  doi     = {10.1109/TCSI.2011.2169889}
}
\end{document}